\documentclass{aastex}
\usepackage{graphicx}
\newcommand{\gae}{\mathrel{>\kern-1.0em\lower0.9ex         
\hbox{$\sim$}}}

\begin{document}        

\title{X-RAY SYNCHROTRON EMITTING FE-RICH EJECTA IN SNR RCW 86}

\author{Jeonghee Rho}

\affil{SIRTF Science Center, California Institute of Technology, 
Pasadena, CA 91125}

\email{rho@ipac.caltech.edu}

\author{Kristy K.~Dyer}

\affil{National Radio Astronomy Observatory, Socorro, NM 87801}

\email{kdyer@nrao.edu}

\author{Kazimierz J.~Borkowski and Stephen P.~Reynolds}

\affil{Department of Physics, North Carolina State
University, Raleigh, NC 27695}

\email{Stephen\_Reynolds@ncsu.edu, kborkow@unity.ncsu.edu}

\begin{abstract}

Supernova remnants may exhibit both thermal and nonthermal X-ray
emission.  In a previous study with ASCA data, we found that the
middle-aged supernova remnant RCW 86 showed evidence for both
processes, and predicted that observations with much higher spatial
resolution would distinguish harder X-rays, which we proposed were
primarily synchrotron emission, from softer, thermal X-rays.  Here we
describe {\it Chandra} observations which amply confirm our
predictions.  Striking differences in the morphology of X-rays below 1
keV and above 2 keV point to a different physical origin.  Hard X-ray
emission is correlated fairly well with the edges of regions of radio
emission, suggesting that these are the locations of shock waves at
which both short-lived X-ray emitting electrons, and longer-lived
radio-emitting electrons, are accelerated.  Soft X-rays are spatially
well-correlated with optical emission from nonradiative shocks, which
are almost certainly portions of the outer blast wave.  These soft
X-rays are well fit with simple thermal plane-shock models.  Harder
X-rays show Fe K$\alpha$ emission and are well described with a
similar soft thermal component, but a much stronger synchrotron
continuum dominating above 2 keV, and a strong Fe K$\alpha$ line.
Quantitative analysis of this line and the surrounding continuum shows
that it cannot be produced by thermal emission from a cosmic-abundance
plasma; the ionization time is too short, as shown both by the low
centroid energy (6.4 keV) and the absence of oxygen lines below 1 keV.
Instead, a model of a plane shock into Fe-rich ejecta, with a
synchrotron continuum, provides a natural explanation.  This requires
that reverse shocks into ejecta be accelerating electrons to energies
of order 50 TeV. We show that maximum energies of this order can be
produced by radiation-limited diffusive shock acceleration at the
reverse shocks.  In an Appendix, we demonstrate that an explanation of
the continuum as due to nonthermal bremsstrahlung is unlikely.

\end{abstract}        


\clearpage       
\section{NONTHERMAL X-RAYS AND RCW 86}

While most shell supernova remnants (SNRs) show thermal X-ray spectra
dominated by emission lines of highly ionized species of C, N, O, Ne,
Mg, Si, S, and Fe, a few show featureless spectra which can be well
understood as synchrotron emission: SN 1006 \citep{Koyama95, dyer01};
G347.3-05 \citep{koyama97, slane99}, G266.2-1.2 \citep{slane01}, and
G28.6-0.1 \citep{bamba01}.
The absence of lines means that a synchrotron interpretation is almost
unavoidable; impossibly low abundances and/or peculiar physical
conditions would be required to suppress lines completely in a thermal
plasma \citep[e.~g.,][]{hss86, laming98}, while nonthermal
bremsstrahlung implies the presence of electrons that should excite
lines as well, and inverse-Compton scattering of any photon population
would result in far too hard a spectrum.  If synchrotron emission can
dominate the spatially-integrated X-ray spectrum of a few remnants, it
is likely to play some role in many more, e.~g., by totally dominating
thermal emission in selected locations within them, or by overwhelming
thermal continua (but not lines) in thermally-emitting regions.  The
signature of an X-ray spectrum with both thermal and synchrotron
radiation would be unusually weak lines, diluted by the synchrotron
continuum.  We interpreted weak X-ray lines in the remnant RCW 86 by
such combination of thermal and synchrotron emission, based on ASCA
observations \citep[hereafter BRRD]{brrd01}. A very similar
interpretation of RCW 86 spectra was also proposed by \citet{bamba00},
based on the ASCA Performance Verification data.

RCW 86 (G315.4--2.3) is a large (42\arcmin\ in diameter) shell-like
SNR with a moderate blast wave speed in the range of 400--900 km
s$^{-1}$, determined from observations of Balmer-dominated shocks
which almost completely encircle the remnant
\citep{lobl90,smith97,ghrs01}. At a kinematic distance of 3 kpc, its
large size and the moderate blast wave speed imply a mature ($\sim
10^4$ yr old) remnant \citep{rosado96}.  The bright complex of optical
emission filaments at its SW corner, to which the optical designation
``RCW86'' actually refers, has been generally interpreted as a region
where the blast wave impacted a dense ($\sim 10$ cm$^{-3}$)
interstellar cloud. This is also where X-ray and radio emission are
the brightest.  Radio observations by the MOST telescope at 843 MHz
\citep{wg96}, with an angular resolution of about $45^{\prime\prime}$,
and by the Australia Telescope Compact Array (ATCA) 
at 1.3 GHz \citep{dsm01}, with
$8^{\prime\prime}$ resolution, show a complete shell of varying
brightness.  The radio morphology in the SW region of the remnant is
somewhat unusual, consisting of a bright ridge of emission mostly
interior to optical and soft X-ray filaments which mark the location
of the blast wave. ASCA Performance Verification X-ray spectra showed
strikingly weak X-ray lines, much weaker than expected from a normal
(solar) abundance plasma, with the exception of the Fe K$\alpha$ line
at 6.4 keV \citep{vink97}. These results were confirmed by deeper ASCA
observations (BRRD) and by more recent BeppoSax observations
\citep{bvfms00}. A substantial Fe K$\alpha$ equivalent width in the SW
suggests enhanced Fe abundance with respect to solar, irrespective of
the origin of the underlying high-energy continuum, which led
\citet{bvfms00} to consider a possibility that we are observing
metal-enriched supernova (SN) ejecta.

In our earlier work (BRRD), we used long ASCA observations to deduce
the presence of three spectral components: a soft thermal component
($kT \sim 0.8$ keV), with a relatively long ionization timescale
($\tau \equiv n_e t \sim 2 \times 10^{11}$ s cm$^{-3}$) and with
approximately solar abundances, modeled as a plane shock (XSPEC model
{\it pshock}); a synchrotron component, modeled as the high-energy
tail of the radio spectrum with XSPEC model {\it srcut}; and a hot
thermal component ($kT \sim 5$ keV, $\tau \sim 5 \times 10^{8}$ s
cm$^{-3}$) required to account for Fe K$\alpha$ emission.  These
components varied in relative strength in three regions in the
southwest corner of RCW 86.  The {\it srcut} model is just the
synchrotron emissivity of a power-law electron distribution with an
exponential cutoff, calculated numerically \citep{rk99}; it is
characterized by a single free parameter, the peak frequency $\nu_{\rm
rolloff}$ radiated by electrons with the $e$-folding energy of the
exponential cutoff.  The fitted values of $\nu_{\rm rolloff}$ varied
from 1 to $4 \times 10^{16}$ Hz, implying (for $B \sim 10 \ \mu$G)
maximum electron energies of order 20 TeV.  The synchrotron fitting
used as inputs the radio fluxes for each region measured from the MOST
image \citep{wg96}, and an assumed constant spectral index of 0.6.

This analysis explained the ASCA data reasonably well, but with ASCA's
spatial resolution of several arcminutes, we were unable to perform
several more rigorous tests of the hypothesis of synchrotron X-ray
emission.  Our spectral fits suggested that the soft thermal emission
is due to nonradiative shocks as identified by H$\alpha$ emission,
while the synchrotron X-rays should show a spatial distribution
similar to radio emission.  In an attempt to verify these predictions,
we observed RCW 86 with the {\it Chandra} X-ray Observatory in
February 2001. We report here on these observations, which not only
confirm our predictions, but allow us to associate synchrotron X-ray
emission with Fe-rich SN ejecta seen through Fe K$\alpha$ line
emission, and identify all high-energy continuum as synchrotron
emission.

\section{{\it CHANDRA} OBSERVATIONS}

\label{chandradata}

RCW~86 was observed for 92 ks February 1-2, 2001 (observation id
500170) by the {\it Chandra X-ray Observatory} \citep{weiss96}
Advanced CCD Imaging Spectrometer \citep[ACIS:][]{garmire92}. Chips
I2, I3, S0, S1, S2 and S3 were used, with the pointing center on
S3. The focal-plane temperature was $-120^{\circ}$ C.  The
observations were corrected for radiation damage (referred to as
charge transfer inefficiency, CTI) that took place early in {\it Chandra's}
mission, following the procedure in \citet{townsley00} which corrects
simultaneously for both position-dependent gain shifts and event grade
changes. The data were extracted and processed using the {\it Chandra}
Interactive Analysis of Observations software package (CIAO) version
2.1.1. The images were corrected with exposure and instrumental
efficiency maps to address chip gaps and mirror vignetting.
Instrumental efficiency maps for each chip were created for each
energy band (0.5 -- 1.0, 1.0 -- 2.0, and 2.0 -- 8.0 keV), for a single
energy in the midrange of each band.  The exposure map has been
correctly weighted for each chip.  The {\it Chandra} images shown here are
binned by a factor of 4 to a resolution of 2\arcsec/pixel.

We analyzed extracted spectra with XSPEC 11.1.0 (for introduction to
XSPEC, see Arnaud 1996\footnote{Or visit the web site at
http://heasarc.gsfc.nasa.gov}). The spectra were grouped by a minimum
of 15 counts per bin.  We have experimented with the source spectra
and background, comparing CTI-corrected spectra, using Pennsylvania
State University software tools (Townsley et al. 2000\footnote{
Available online at http://www.astro.psu.edu/users/townsley/cti/}), and
non-CTI-corrected spectra, using standard CXC CIAO tools.  The
CTI-corrected spectra give more accurate centroids of lines such as Mg
and Si, resulting in a particularly significant improvement for the
front-illuminated chips. The differences are less than 40 eV for the
back-illuminated chip, and although this is less than the ACIS
back-illuminated spectral resolution of 100 -- 150 eV, the CTI
correction provides a significant improvement even for this chip.

The issue of background was not straightforward for this observation
since the southeast corner of the supernova remnant nearly fills the
S3 chip.  Using background from an off-source area and from blank-sky
files did not change the fit results (see Table 1) within the errors.
This is partially due to the small contribution (less than 0.5\%) of
the background to the source counts for RCW 86. Since standard blank
sky background is not compatible with the CTI-corrected spectra, we
have chosen to take background spectra from off-source areas on the
same chip. For the front-illuminated I3 chip with more substantial CTI
effects, varying CTI corrections for different regions of the chip
constrained us to choose for background a part of the same CTI region
as used for the signal, rows 641-768.  While these regions likely
contain source as well as background photons, since they are from the
same chip and have been processed identically with the source regions
they should not introduce large errors into the source
spectra. Residual instrumental background features such as a faint Si
K$\alpha$ line are present in at least one of our
background-subtracted spectra, but since they are very weak they do
not pose any problems for spectral fitting.

\section{RESULTS}

\subsection{Images}

Figure~\ref{fig:totx} shows the image from the I2, I3, and S0--S3
chips of the SW corner of RCW 86, binned to a resolution of 2\arcsec\
and including the full energy range.  The extensive filamentary
structure is quite striking and persists down to the scale of the
resolution.  Most of this structure is parallel to the presumed edge
of the remnant.  However, some filaments can be seen that are almost
radial in orientation, and some emission is considerably more diffuse.
Compared to the ROSAT HRI image \citep{bvfms00,dsm01},
Figure~\ref{fig:totx} shows emission extending considerably further
ahead of the bright X-ray filaments in the S and SW.  Some of that
emission is bounded by a sharp-edged filament, but some (in the
extreme SW corner, for instance), is not.

Figure~\ref{fig:truecolor} contrasts the X-rays in soft (0.5 -- 1 keV;
red) medium (1 -- 2 keV, green), and hard (2 -- 8 keV, blue) energy
bands.  (The soft and hard band images are shown separately in
Figures~\ref{fig:softx} and \ref{fig:hardx}, respectively.)  The
brightest emission is clearly relatively soft with some even softer
emission further from the remnant center.  The network of soft
filaments is remarkably similar to the H$\alpha$ image
\citep[Fig.~\ref{fig:Halpha},][]{smith97}.  The hard X-rays have a
dramatically different morphology, with completely orthogonal
filaments in some locations on S3 and more diffuse emission in I3.
The brightest region in the hard X-ray band is the somewhat localized
knot near the center of the I3 chip.  These morphological differences
strongly suggest a different origin for the hard X-rays than for the
soft and medium bands, a conclusion confirmed by spectral analysis
(see below).

Radio and X-ray morphologies are compared in
Figure~\ref{fig:xrayradioc}, where X-rays below 2 keV are indicated
with green, above 2 keV with blue, as in Figure~\ref{fig:truecolor},
and radio (the Australia Telescope image: Dickel et al. 2001) in red.
While neither X-ray band shows a structure closely resembling radio,
there are clear correlations between the hard X-rays and radio: hard
X-rays often appear at the edges of radio-emitting regions, most
strikingly in a N-S filament of hard X-rays with radio extending
beyond to the east, in the center of the S3 chip.  We have extracted
radio and X-ray profiles from the three rectangular regions shown in
Fig.~\ref{fig:extract}, where we have averaged over the narrow
dimensions of the rectangles.  The profiles are compared in
Fig.~\ref{fig:profiles}.  All three regions show that thin X-ray
filaments are associated with rises (or drops) in radio emission,
including the hard knot in the center of I3. We interpret these as
examples of shocks seen mostly edge-on, with the downstream direction
the one in which radio emission rises.  We shall argue below that
these coincidences support a synchrotron origin for the hard X-ray
emission.

\subsection{Spectra}

Figure~\ref{fig:extract} shows extraction regions for the nonradiative
Balmer filament, for the entire hard region on the S3 chip, for the
synchrotron filaments alone, and for one region on the I3 chip.
The spectrum of the nonradiative shock, visible in optical as a bright
Balmer filament, is soft (Figure~\ref{fig:soft}), without any
measurable emission above 4 keV. In particular, there is no Fe
K$\alpha$ line emission associated with this shock. We modeled its
spectrum with plane shock models {\it pshock} and {\it vpshock} which
are available in XSPEC (these models, described by Borkowski, Lyerly,
\& Reynolds 2001, are based on updated thermal plasma code of Hamilton
\& Sarazin 1984a).  Because of calibration problems for the ACIS S3
detector at photon energies below 0.5 keV, we ignored data below this
photon energy threshold in all our fits.

We also excluded channels with energies from 1.17 keV to 1.31 keV in
our fits to the X-ray spectrum of the soft region. This is the
spectral region with ``missing'' Fe-L shell lines
\citep{bretal00}. Modeling of the Fe L-shell complex in the
equilibrium-ionization MEKAL and all non-equilibrium ionization (NEI)
models in XSPEC is based on calculations by \citet{liedahl}, which did
not include electron excitations to levels higher than $n =
5$. \citet{bretal00} found that Fe L-shell lines originating at these
highly excited levels cannot be neglected in the ASCA spectrum of
Capella. In Capella these lines are produced by $6 \times 10^6$ K (0.5
keV) plasma in collisional ionization equilibrium, and they cluster
near 1.2 keV. Because plasma temperatures are similar in Capella and
in our soft regions associated with Balmer filaments, we excluded
channels near this energy where we expect Fe L-shell lines originating
at $n > 5$ which are missing in the NEI spectral code.

Results of plane shock fits to the soft spectrum are tabulated in
Table~\ref{tab:models}. In the first fit with {\it wabs $+$ pshock}
models, relative abundances of heavy elements are fixed at solar
values, but their absolute abundance with respect to H and He is
allowed to vary. In the second fit with {\it wabs $+$ vpshock} models,
abundances of individual elements are allowed to vary independently
with the following exceptions: the C abundance is tied to the N
abundance, and Ca and Ni are tied to Fe. The second fit reproduces the
data reasonably well (reduced $\chi^2 = 2.7$, see
Figure~\ref{fig:soft}). Both models underpredict emission at
energies below 0.5 keV by up to a factor of 2, which may be either
caused by calibration/instrumental problems or by the presence of
additional gas with much lower temperatures in the spectral extraction
region. Electron temperatures are relatively low, 0.41 -- 0.43 keV,
shock ionization ages are relatively short, $(3-5) \times 10^{10}$
cm$^{-3}$ s, and elemental abundances are close to solar. The largest
departures from solar abundances occur for Fe and Mg, whose apparent
subsolar abundances might be caused by moderate depletion onto dust
grains and by overlapping Fe L-shell lines from high $n$ levels, which
are adjacent to the Mg K$\alpha$ line. An apparent excess of emission
near 3 keV hints at the presence of the Ar K$\alpha$ line (Ar is not
included in the NEI code).

There is no evidence for the presence of excess emission at high
energies, suggesting that thermal plane shock models with constant
electron temperature and approximately solar abundances provide a
satisfactory description of the soft spectrum. This finding fully
confirms our previous inferences (BRRD) about the thermal nature of
the soft X-ray emitting filaments which are associated with the blast
wave propagating into the ambient ISM.

We realize that our very simple single plane-shock model for the soft
emission is quite unrealistic.  In the multiband X-ray image
(Figure~\ref{fig:truecolor}), it is clear that the bright thermal
filaments have spectral differences on spatial scales much smaller than
the size of our spectral extraction regions, and extending down nearly
to the {\it Chandra} spatial resolution limit.  The small values of ionization
timescale $\tau$ we find may be artifacts of using
constant-temperature shock models to fit regions with clearly varying
temperatures.  The focus of the present investigation is understanding
the harder emission, and for that purpose we simply require an
adequate description, if not a full explanation, of the soft emission.
We defer to a later work a more extensive exploration of the nature of
the soft emission and possible explanations for the short ionization
timescale.

Spectra of hard regions are shown in Figure~\ref{fig:hard} (the entire
synchrotron-emitting region in the NE corner of the S3 chip) and
Figure~\ref{fig:synchrotronI} (synchrotron filaments on S3 and I3
chips).  (See Figure~\ref{fig:extract} for the locations of these
regions.)  Results of fits with the {\it srcut} + low-temperature {\it
vpshock} models are shown in these Figures and tabulated in
Table~\ref{tab:models}. The Fe K$\alpha$ line is modeled by an
additional pure Fe {\it vpshock} model (see \S~\ref{ejecta} for
further details). Ideally, one uses the {\it srcut} model with radio
information as input: the 1 GHz radio flux of the region and the
spectral index.  The integrated spectral index of the entire RCW 86
remnant is about 0.6, based on single-dish observations (typical
errors are $\pm 0.05$), but accurate fluxes cannot be determined
because the only available radio images are interferometer images, in
which the absence of very short antenna baselines means that very
smoothly distributed flux is resolved out.  In fact, the MOST image at
843 MHz \citep{wg96} contains less than 40\% of the single-dish flux
at that frequency, while the ATCA image \citep{dsm01} contains about
70\% of the 1.3 GHz single-dish flux.  In local regions, the total
flux can be in error by even larger factors.  As a result, we cannot
use the measured radio fluxes to constrain the X-ray modeling.
Instead, we have fixed the radio spectral index at 0.6 and actually
fit for the radio flux.  The values we obtain are quite reasonable and
within factors of 2 of the fluxes measured from the radio maps (which
differ between themselves by a comparable factor).

Nonthermal X-ray emission from RCW 86 was reported from {\it Rossi}
X-ray Timing Explorer (RXTE) observations by \citet{allen99}.  The
spatial resolution of RXTE is only about one degree, so it cannot
resolve RCW 86.  \citet{allen99} report that its spectrum observed
with the PCA instrument is well described by a power-law with energy
index $\Gamma = 3.3 \pm 0.2$ between 10 and 24 keV.  Our {\it srcut}
model, with parameters fit to the hard region as described above, is
slightly curved in that interval, with a mean slope between 10 and 20
keV of $\Gamma = 3.5$.  It predicts a flux at 10 keV smaller than that
quoted by \citet{allen99} by a factor of 12.5, and gives a 1 GHz radio
flux of 2.59 Jy.  If we simply scale up our X-ray model to the flux
level reported by \citet{allen99} by that factor, the predicted radio
flux at 1 GHz is 32 Jy.  Green (2001\footnote{Available online at
http://www.mrao.cam.ac.uk/surveys/snrs/}) reports the observed 1 GHz
flux of RCW 86 to be 49 Jy. We regard this crude agreement, to within
50\% after an extrapolation of nine orders of magnitude in photon
frequency, as evidence of broad consistency between the whole-remnant
spectrum seen by RXTE and the local spectra we fit in individual
synchrotron filaments, consistency which represents strong support for
the idea of a substantial synchrotron component in the spectrum of RCW
86.

For the soft thermal regions, we also attempted to bound any
nonthermal emission.  We obtained a quantitative upper limit on the
presence of a synchrotron component by adding to the fit an {\it
srcut} component, with a 1 GHz flux determined from the Australia
Telescope observations \citep{dsm01} and assuming a spectral index of
0.6, allowing $\nu_{\rm rolloff}$ to vary.  The fits rejected values
of $\nu_{\rm rolloff}$ larger than about $2 \times 10^{16}$ Hz.

Attempts to fit the hard continuum and the Fe K$\alpha$ line with pure
thermal models are discussed below; we find that the resulting
extremely short ionization timescales are inconsistent with the known
age of the remnant, which should be the shock age if the Fe K$\alpha$
line is due to solar abundance Fe from the ISM.

\section{DISCUSSION}

\subsection{Morphological Evidence for Synchrotron Emission}

We believe that the striking difference in morphology of hard and soft
X-rays is strong evidence in favor of the synchrotron interpretation
of the bulk of the harder X-ray continuum (above 2 keV).  
The spectrum of regions bright in X-rays below 2 keV
is well described by thermal plane-shock models.  The hard X-rays show
a dramatically different structure, often occurring on the edges of
regions of radio emission.  It is sometimes claimed that a nonthermal
interpretation of X-rays requires that the X-ray and radio
morphologies be identical.  This is only true to the extent that
synchrotron losses on electrons can be neglected, rarely the case
except for very young remnants such as the historical Galactic
remnants.  In a magnetic field of 10 $\mu$Gauss, electrons radiating
the peak of their synchrotron emission at $h\nu = 4$ keV lose half
their energy in 900 years, far less than the presumed age of order $10^4$
yr, while electrons radiating chiefly at 1 GHz will last for over
$10^7$ yr.  Thus we should expect synchrotron X-rays to be seen near
the shocks where the electrons are presumably accelerated, while radio
emission can be seen there as well as far downstream.

In the complex inhomogeneous environment of RCW 86, shock structures
can be found in various locations inside the outermost X-ray and radio
emission.  Some may be parts of the reverse shock or secondary shocks
driven into denser clouds, and projection effects further complicate
the interpretation.  However, a particularly clear example of what
appears to be an internal shock seen edge on is indicated in the
central (``hard-filament'') extraction region on chip S3 in
Figure~\ref{fig:extract}, and profiles along the indicated direction
are shown in Figure~\ref{fig:profiles}.  Here, both radio and X-ray
emission appear fairly abruptly; the X-ray emitting region is only a
few tens of arcseconds wide, while radio emission persists much further,
just as expected if this is an outward facing shock at which electrons
are newly accelerated.  At a distance of 3 kpc, $1^{\prime\prime}$
corresponds to $0.015$ pc or $4.5 \times 10^{16}$ cm.  Shock
velocities from optical or X-ray analysis are of order 600--900 km
s$^{-1}$, so that one might expect post-shock flow velocities in the
plane of the sky to be of the same order, but somewhat smaller.  In 900
years, a flow at 500 km s$^{-1}$ travels a distance $l = 0.46$ pc
corresponding to an angular distance of $32^{\prime\prime}$.  This
compares reasonably well with the thickness of the overall X-ray
emitting peaks in the profiles of Figure~\ref{fig:profiles}, allowing
for projection effects and effects of magnetic-field geometry, which
can account for the irregular shapes of the profiles.  In general, we
should expect synchrotron-emitting regions to have thicknesses of
order a few tens of arcseconds, though if a shock is seen close to
face-on, one will obtain more diffuse, fainter extended emission.
These properties do appear to characterize the structure of the hard
X-ray emission as well as can be expected.

We do not see clear evidence for hard X-ray emission from the extreme
outer edge of the X-ray emission seen in the lower part of the S3
chip, where at least one part of the blast wave is presumably
encountering undisturbed upstream material.  However, the overall
faintness of the emission (presumably mostly thermal) is such that a
faint nonthermal component might be hard to discern.  Unfortunately,
the emission there is too faint for useful spectral analysis with
these data.  Interior shocks could be portions of the outer blast wave
encountering denser material and seen in projection, or portions of a
reverse shock moving back into the interior from such an encounter.
It seems that such interior shocks are where most hard X-rays
originate, though not at the H$\alpha$-bright nonradiative shocks
which are probably the blast wave moving into partially neutral
material.  This is an extremely interesting result, as it suggests
that electrons may be accelerated at reverse shocks as well as at
forward shocks -- perhaps preferentially at reverse shocks in RCW 86.
In the prototype synchrotron-X-ray-dominated remnant, SN 1006, no
obvious reverse-shock structure is seen, and the hardest emission
appears to come from closest to the outer remnant edge.  While a
reverse shock is almost certainly present in SN 1006, one can explain
its absence in radio or X-ray nonthermal emission if the magnetic
field threading the ejecta is very weak, as one might expect if it is
just the magnetic field of the progenitor, enormously diluted by
expansion with flux freezing.  The situation appears to be different
in RCW 86.

\subsection{Spectroscopic Evidence for Fe-rich SN ejecta} 

\label{ejecta}

The Fe K$\alpha$ line at 6.4 keV in RCW 86 has been puzzling since
\citet{vink97} reliably determined its energy, because such a low
energy implies the presence of only very low ionization stages of Fe
in the X-ray emitting plasma.  \citet{vink97} suggested strong
deviations from a Maxwellian electron distribution in order to account
for both the presence of the Fe line and the weakness of other
emission lines at low energies. Subsequent studies
\citep[BRRD;][]{bamba00,bvfms00} have generally postulated its origin
in a hot ($\sim 5$ keV) plasma in the remnant's interior. A
substantial Fe K$\alpha$ strength suggests that Fe is overabundant by
a factor of several with respect to cosmic (solar) abundances, but
this finding is somewhat uncertain in view of the poor photon
statistics and uncertainties in atomic data. Although our {\it
Chandra} observations of RCW86 also suffer from poor photon statistics
at high energies, the high spatial resolution of {\it Chandra} allows
us to spatially separate the synchrotron-emitting regions and the soft
thermally-emitting X-ray filaments. We find that Fe K$\alpha$ is
associated with the synchrotron-emitting regions, and not with the
thermal filaments. Apparently, fast electrons capable of exciting Fe
K$\alpha$ are present in synchrotron emitting regions, and absent in
soft X-ray filaments.

It seems natural to assume that the Fe K$\alpha$ line is produced in a
hot thermal plasma with an approximately cosmic composition, where
most free electrons are provided by the ionization of H and He. In
this hypothesis the observed high-energy continuum must be thermal,
because a substantial contribution from synchrotron emission would
demand a high derived Fe abundance for the line to rise above the
continuum, and this would be obviously in conflict with the assumed
solar composition.  This interpretation unexpectedly encounters a
timescale problem which makes it untenable, as we now demonstrate.  We
use the fitted ionization timescale $n_e t$, and derive $n_e$ from the
observed X-ray flux in the model to obtain a characteristic age of the
shocked plasma under this assumption.

For a plasma consisting mostly of H and He, we can set a lower limit
to the electron density in the synchrotron-emitting region by fitting
a thermal NEI model to the high-energy {\it Chandra} spectrum.  But
because the {\it Chandra} spectrum is rather noisy in the vicinity of
the Fe K$\alpha$ line, we shall obtain the temperature, Fe abundance,
and ionization timescale from high-quality ASCA GIS spectra presented
by us previously (BRRD). (The analysis of the Fe K$\alpha$ line in
BRRD was influenced by an error in all NEI models in XSPEC v10, which
resulted in a missing Fe K$\alpha$ fluorescence line at extremely low
ionization timescales relevant for RCW86.  This error is not present
in the NEI models in XSPEC v11.) The corrected values for the {\it
vnei} model fit to GIS spectra above 5.5 keV are: plasma temperature
of 5.0 keV (with a 90\%\ confidence interval of 3.5--7.6 keV), a very
low ionization timescale $\tau$ of $2.8 \times 10^8$ cm$^{-3}$ s (less
than $6.6 \times 10^9$ cm$^{-3}$ s with 90\%\ confidence), and an Fe
abundance of 1.6 solar (with 90\%\ confidence interval 1.1--2.5). By
using these best-fit values, we find that this model must be
normalized to 0.0048 (in XSPEC units for thermal models: normalization
= $(10^{-14}/4\pi D^2) \int n^2 dV$, where $D$ is the distance to the
source in cm and $n$ is the electron density).  This value is required
to account for the high energy emission seen in Figure
\ref{fig:hard}, implying an emission measure $EM = \int n_e n_H dV$ of
$5.2 \times 10^{56} d_3^2$ cm$^{-3}$ (where $d_3$ is the remnant's
distance in units of 3 kpc). The extraction region is approximately
$2\farcm4 \times 5\farcm0$ in size, spanning the entire width of the
synchrotron-emitting region. This translates into a surface area of
$2.1 \times 4.2$ pc$^2$ at 3 kpc distance. Assuming 4.2 pc depth, we
estimate the volume of the X-ray emitting region to be $37 d_3^3$
pc$^3$. For H and He dominated plasma, we then arrive at an electron
density $n_e$ of $0.76 f^{-1/2} d_3^{-1/2}$ cm$^{-3}$, where $f$ is
the volume filling fraction of the hot gas. The plasma characteristic
``age'' is then $\le \tau/n_e = 12 f^{1/2} d_3^{1/2} $ yr, which is
far too short compared with the age of the remnant.

The upper limit on $\tau$ from fits to ASCA GIS data is of course much
larger than the best-fit value of $2.8 \times 10^8$ cm$^{-3}$ s used
in these estimates, but we can establish a more stringent upper limit
on $\tau$ which applies for plasmas with approximately cosmic
composition. We noted previously an extreme sensitivity of NEI models
to ionization timescale (Figure 8 in BRRD). At extremely low
ionization timescales, most abundant elements such as O have not been
stripped of their L-shell electrons, resulting in weak X-ray lines in
the {\it Chandra} ACIS spectral range and low X-ray fluxes at low
energies. As the ionization timescale increases, electrons are
stripped from their L shells, and the soft X-ray flux increases
dramatically.  By varying the ionization timescale, we find that the
{\it vnei} model with the temperature and Fe abundance quoted above
and $N_H = 4.2 \times 10^{21}$ cm$^{-2}$ (from model B in
Table~\ref{tab:models}) produces strong O lines at low energies which
are inconsistent with {\it Chandra} spectra for $\tau > 1 \times 10^9$
cm$^{-3}$ s. (We obtained this conservative upper limit by requiring
that the calculated spectrum not exceed the observed count rate
at the location of O lines. But in contrast to observations, O lines
are very prominent in strongly-underionized models, so based on
consideration of the spectral shape one can probably reject models
with $\tau$ as short as $5 \times 10^8$ cm$^{-3}$ s.)  This constraint
on ionization timescale gives an age of less than 40 yr. By assuming a
much higher plasma temperature we can perhaps push this limit to
$\sim100$ yr, which is still much less than expected for a relatively
old remnant such as RCW 86. This timescale problem is not an entirely
new finding, as an inspection of previous results obtained with
various satellites demonstrates it clearly, but new constraints from
{\it Chandra} are particularly severe.  The Fe K$\alpha$ line and
continuum at high energies cannot be easily produced by a thermal plasma
with approximately solar abundances.

A consideration of Coulomb interactions between ions and electrons
(see the Appendix) leads us to a conclusion that the observed
continuum in regions of the remnant with hard spectra cannot be
produced by nonthermal bremsstrahlung in cosmic abundance
plasmas. This leaves synchrotron radiation as the preferred
explanation for the observed continuum in the whole {\it Chandra}
spectral range, including high energies. But the Fe K$\alpha$ line
must be produced by collisional excitations and ionizations by much
less energetic thermal electrons. Because the Fe K$\alpha$ line
strength relative to the high energy continuum already implies above
solar abundances, it is likely that we are observing highly-enriched
SN ejecta. SN ejecta were already considered in some detail by
\citet{bvfms00}, but now we have indirect evidence for the presence of
very Fe-rich ejecta.  For simplicity, we considered pure Fe-rich
ejecta and modeled them by the {\it vpshock} model with temperature 5
keV and ionization timescale $10^9$ cm$^{-3}$ s. These plasma
parameters are just rough guesses: at much lower temperatures most
thermal electrons are not capable of exciting the Fe K$\alpha$ line,
while at ionization ages above a few $\times 10^9$ cm$^{-3}$ s, Fe L
shell emission becomes very prominent (in conflict with {\it Chandra}
observations). In addition to the pure Fe {\it vpshock} model which
accounts for the Fe K$\alpha$ line, we used the {\it srcut} model for
modeling synchrotron emission. The blast wave seen in projection
against synchrotron-dominated regions is modeled by a low-temperature
{\it vpshock} model, either with solar abundances, or with abundances
determined from our fit to the spectrum of the Balmer-dominated
nonradiative shock. Results are tabulated in
Table~\ref{tab:models}.

The Fe K$\alpha$ line is unambiguously detected only in the spectrum
of the entire synchrotron-emitting region on the NE corner of the S3
chip (see Figure~\ref{fig:extract}).  The emission measure $EM_{Fe}
= \int n_e n_{Fe} dV$ of pure Fe-rich ejecta is $3.1 \times 10^{52}
d_3^2$ cm$^{-3}$, based on the normalization of model B in
Table~\ref{tab:models}.  With the estimated X-ray emitting volume $V$
of $37 d_3^3$ pc$^3$, we arrive at $n_e = 0.017 (n_e/10n_{Fe})^{1/2}
f^{-1/2} d_3^{-1/2}$ cm$^{-3}$. The factor $(n_e/10n_{Fe})^{1/2}$ is
of the order of unity for a pure Fe plasma with $\tau = 10^9$ cm$^{-3}$
s, but it is somewhat uncertain because of the poorly known electron
distribution function and a possible contribution to $n_e$ from other
heavy elements which might be present in the ejecta. The electron
distribution function may well be non-Maxwellian in such a low-density
and high-temperature plasma because of the long Coulomb collision
timescale. In particular, the distribution may be bi-modal, with free
preshock electrons heated to high temperatures at the collisionless
shock, and with much colder electrons released in the process of
collisional ionization in the postshock region \citep{hamsar84b}. The
electron density $n_e$ just quoted refers to the hot component, and
not to cold electrons with energies much below the Fe K-shell
ionization threshold.  For a pure Fe plasma with electron density
0.017 cm$^{-3}$ and $\tau = 10^9$ cm$^{-3}$ s, the plasma age is about
2000 yr, which is consistent with the relatively large age of RCW 86.

The presence of low-density Fe-rich ejecta in RCW 86 is consistent
with the large observed equivalent widths of the Fe K$\alpha$ line,
which already led \citet{bvfms00} to suggest the possible presence of
SN ejecta in this remnant. The morphology of synchrotron-emitting
filaments apparently associated with the ejecta is also
suggestive. The X-ray and radio synchrotron-emitting filaments are
located mostly interior to the bright nonradiative and radiative
shocks in the ``knee'' region of the remnant.  These filaments form a
ridge which extends much further to the east, and which is clearly
located well within the blast wave outlined by nonradiative
Balmer-emitting shocks \citep{smith97}. The brightest part of this
ridge is in the region of highest pressures where the most intense
interaction of supernova ejecta with the ambient ISM is occurring, as
judged by the presence of bright radiative and nonradiative shocks.
This morphology is consistent with identification of this ridge with
the reverse shock. In addition to solving the short timescale puzzle,
the presence of low-density Fe-rich SN ejecta also removes a need for
a mixed thermal-nonthermal continuum postulated by us previously for
the synchrotron emission region \citep[BRRD; see
also][]{bamba00}. This required a coincidental match in intensities of
synchrotron and thermal continuum emission in the 1--10 keV
range. This is no longer necessary, since the observed continuum is
mostly synchrotron emission, and the thermal contribution to the
continuum may be as much as 2 or 3 orders of magnitude fainter.

\subsection{X-Ray Synchrotron Spectra}

No synchrotron component was required to obtain a reasonable fit for
the soft region (Fig.~\ref{fig:soft}); an upper limit to the rolloff
frequency of about $2 \times 10^{16}$ Hz was obtained as described in
the previous section.  We find values of $\nu_{\rm rolloff}$ of $(7 -
10) \times 10^{16}$ Hz from the fits to hard regions, with remarkably
little dispersion.  These values are two to three times higher than
those we reported in BRRD, for two reasons.  First, the enormously
improved spatial resolution of {\it Chandra} allows us to select spectral
regions for extraction that are localized to the very hardest X-ray
emission, whereas the fits based on ASCA data are necessarily averages
over much larger regions, lowering the mean value of $\nu_{\rm
rolloff}$.  Second, as described above, we now believe the entire hard
continuum, rather than just part of it, to be due to synchrotron
X-rays.

The values of $\nu_{\rm rolloff}$ can be understood in the framework
of simple estimates of particle shock-acceleration times (see, for example,
Reynolds 1998).  If the maximum energy that electrons can reach is
limited by radiative losses, the energy at which energy gains by
acceleration equal losses due to radiation is about

$$E_m \sim 50 \left(\eta R_J \right)^{-1/2} 
      \left(B_1/3 \ \mu{\rm G}\right)^{-1/2} u_8 \ {\rm ergs,}$$
where $\eta$ is the ratio of electron mean free path to gyroradius
(and is assumed constant), $B_1$ is the upstream magnetic-field
strength, and $u_8$ is the shock speed in units of $10^8$ cm
s$^{-1}$. We have assumed a shock compression $r$ of 4.  The factor
$R_J$ is Jokipii's (1987) expression for the obliquity-dependence of
the particle acceleration rate; where the shock is nearly
perpendicular (i.~e., the shock normal nearly perpendicular to the
upstream magnetic field direction), 
$$R_J \sim \left(2 \over {1 + r}\right) \left( 1 + \eta^2 \right)^{-1}$$
so that for $\eta$ substantially greater than 1, $\eta R_J \propto
\eta^{-1}$ and $E_m \propto \eta^{+1/2}$.  Electrons of this energy
radiating in the downstream field $B_2$ produce their peak power at a
frequency
$$\nu_c \sim 5 \times 10^{16} \left( \eta R_J \right)^{-1} 
\left(B_2/4 \ B_1\right) {u_8}^2.$$
If reverse shock velocities are comparable to the 
blast-wave velocities of 400 -- 900 km s$^{-1}$, then, we
can reach the observed values of $\nu_{\rm rolloff}$ for compression
ratios substantially larger than 4, and/or values of $\eta$
substantially greater than 1.  (Notice that the characteristic
frequency radiated by electrons whose maximum energy is set by
radiative losses is independent of the magnetic-field strength.)  It
should not be surprising that the hard emission is so restricted in
spatial distribution, then; many of the shocks present in RCW 86 will
not be fast enough to produce synchrotron X-ray-emitting electrons.
Only in a few regions where the conditions are right will we expect
such electrons to be accelerated.  

We should add that the model we use to describe the synchrotron
emission, {\sl srcut}, is highly oversimplified: it is just the
synchrotron emissivity of a power-law distribution of electrons with
an exponential cutoff (the fastest plausible cutoff for shock-accelerated
particles; see references in  Reynolds \& Keohane 1999),
radiating in a uniform magnetic field. Much
more elaborate models of spherical remnants in which radiative losses
limit electron acceleration were described in \citet{R98}; those
models, not available in XSPEC, produced synchrotron spectra with a
somewhat different shape than {\it srcut}.  While the curvature within
the {\it Chandra} bandpass is not enough to distinguish the
difference, the inferred values of $\nu_{\rm rolloff}$ would probably
be slightly different for a true loss-limited model.  In fact, since
any spatial inhomogeneities will broaden the cutoff, {\it srcut}
represents the steepest possible cutoff of emission, and will give the
highest rolloff frequency compatible with the observed data.  Any more
realistic model should produce a somewhat lower value.

Our fits to the synchrotron component could be substantially better
constrained if radio fluxes were known more accurately.  The most
straightforward way to accomplish this would be to map RCW 86 with
a large single dish, such as the Parkes telescope, and combine those
data with the ATCA interferometer image.  This would allow considerably
firmer determinations of $\nu_{\rm rolloff}.$

We have argued that the hard continuum and Fe K$\alpha$ line tend to
be found together, strongly suggesting that it is the reverse shock
that is responsible for accelerating particles to the highest
energies.  This conclusion is important to confirm, since it seems not
to be obviously the case in other remnants with synchrotron X-ray
emission.  There is no obvious reason that reverse shocks should not
be as effective as the blast wave at accelerating particles, for the
same values of physical parameters such as shock speed and $\eta.$ In
RCW 86, blast-wave emission, while detectable, is so faint that
synchrotron X-rays would be below the level of background, so we
cannot rule out electron acceleration to comparably high energies at
the blast wave.

Observing particle acceleration at the reverse shock may also have
interesting consequences for the acceleration of cosmic rays.  The
direct acceleration of enriched material in supernova remnants might
play a role in models of Galactic cosmic-ray acceleration and of
chemical enrichment of the light elements Li, Be, B in the early
Galaxy, when the carbon and oxygen from which these elements are
spalled were less abundant in the general interstellar medium.

Even more intriguing is the possibility that the reason RCW 86 shows
particular evidence for particle acceleration at the reverse shock is
that the radiating particles are not electrons but positrons produced
by decay of radioactive $^{56}$Ni synthesized in the explosion.  A
longstanding theoretical difficulty in electron acceleration has been
the ``injection problem'': thermal electrons have such small gyroradii
that they see the shock as a smooth transition and not the
discontinuity in flow speeds required for standard diffusive shock
acceleration 
%\citep[see, e.g.,][for extensive discussion]{levinson92,levinson94}.
(see, e.~g., Levinson 1992, 1994 for extensive
discussion).  
Various suggestions have been made about how Nature
evidently overcomes this difficulty.  However, it has been noted
\citep{rl79,ejr89}
that positrons released by radioactive decay of $^{56}$Ni and
$^{56}$Co have typical energies of order 1 MeV, where they should see
the shock as a discontinuity and be accelerated as efficiently as
protons of similar energies.  A reverse shock moving into Fe-rich
ejecta should encounter a population of ``pre-injected'' energetic
positrons which can be immediately accelerated efficiently.  While
this is a possibility difficult to rule out, we note that the decay
lifetimes of $^{56}$Ni and $^{56}$Co are both less than a year; such
positrons were almost entirely produced about $10^4$ years ago, just
after the supernova. Any X-ray-synchrotron-emitting positrons cannot
have been accelerated soon after that, because the radiative lifetime
is less than 1000 yr.  However, 1 MeV positrons will lose energy
primarily by adiabatic expansion losses, which could reduce them to no
better than thermal electrons after $10^4$ yr.  This fascinating
possibility must remain a speculation at this time.

\section{CONCLUSIONS}

The major conclusion of our paper is that the detailed morphology of
soft and hard X-rays in RCW 86 strikingly supports the hypothesis of 
different origins.  Our spectral analysis confirms
that the hard X-rays are best described as a synchrotron continuum
with Fe K$\alpha$ emission that must come from a thermal component
with a substantial overabundance of iron with respect to solar.  This
result implies that some of the original supernova ejecta are still
unmixed after $\sim 10^4$ yr, and that the reverse shock into those
ejecta can also accelerate electrons to X-ray synchrotron-emitting
energies of order 50 TeV.

Both these conclusions substantially complicate the task of
understanding the X-ray spectra of middle-aged supernova remnants.
Apparently even after ages of order $10^4$ years, remnants are not in
the simple Sedov phase, and the possibility of synchrotron X-ray
emission confusing the analysis of thermal emission cannot be
neglected.  The only ultimate solution to this problem requires X-ray
spectroscopy with much higher energy resolution.  (Grating resolution
is adequate, but grating observations of large extended sources with
{\it Chandra} or {\it XMM-Newton} are
virtually uninterpretable.)  The {\sl Astro-E2} microcalorimeters hold
out the possibility of spatially resolved spectroscopy with enough
resolution to allow the inference of temperatures from line complexes
alone without requiring any assumptions about the continuum.  In that
case, one can immediately determine the extent of synchrotron
contributions to the continuum.

With {\it Chandra's} spatial resolution, we can for the first time
discern discrete shock features and minimize projection effects, which
should aid substantially in the detailed thermal modeling that will be
the next step in understanding RCW 86.  The hard X-ray emission
appears to delineate particular shocks, where both radio and X-ray
emitting electrons are accelerated.  The greater extent of radio
emission behind such features is expected from preferential energy
losses on the much more energetic X-ray emitting electrons.  The width
of the X-ray emission is quite consistent with this interpretation,
for expected post-shock velocities of hundreds of km s$^{-1}$ and
magnetic field strengths of order 10 $\mu$Gauss.

The observed rolloff frequencies in areas containing synchrotron
continua can be understood in terms of radiative-loss-limited shock
acceleration, but simple estimates require favorable conditions to
reach the inferred electron energies.  In particular, nearly
perpendicular shocks reach higher energies more readily, and this may
be one factor that distinguishes shocks that produce synchrotron X-ray
continua from those which do not.  We repeat, however, that the {\it
srcut} model is highly simplified, and fits with more elaborate models
might produce somewhat different values for $\nu_{\rm rolloff}$.

Progress in understanding RCW 86, and the general problems it raises
for the understanding of supernova remnants, requires further
observations in various bands.  {\it XMM-Newton} observations can
obtain significantly higher signal-to-noise ratios at high energies,
allowing testing in other regions of our claim of an association of Fe
K$\alpha$ emission with regions of hard continuum.  Optical and
infrared observations in principle should be able to detect
synchrotron continuum at a level exactly predictable from our fits,
but distinguishing it from the confusing emission might prove
extremely challenging.  Since our fitted values of $\nu_{\rm rolloff}$
are far above optical frequencies, we expect that there should be
little difference between the optical or IR morphology of synchrotron
emission and the radio morphology.  A strategy to search for extremely
faint diffuse emission then might center on sharp features in the
radio image, such as the edge coincident with a hard X-ray filament in
the middle of chip S3.

\acknowledgements 

Support for this work was provided by NASA through {\it Chandra}
grant G01-2077A and G01-2077B. 
J. R.  acknowledge the support 
of California Institute of Technology, the Jet Propulsion Laboratory,
which is operated under contract with NASA.
Kristy Dyer would like to acknowlege work supported by the National
Science Foundation under Grant No. 0103879.
We thank Sara Gallagher and Leisa Townsley for help with CTI correction of
{\it Chandra} RCW 86 data, John Dickel for making Australia Telescope
radio data available to us, and Chris Smith for his RCW 86
H$\alpha$ image. We acknowledge discussions with Don Ellison about particle 
acceleration in reverse shocks, including acceleration of positrons 
produced in radioactive decays. 

\appendix 

\section{Appendix}

Here we demonstrate that nonthermal bremsstrahlung is unlikely to be
important in the regions we have analyzed in RCW 86.  First, we shall
show that such an explanation for the lower-energy continuum would
require a huge unseen cold-electron population of which the observed
$\sim 2$ keV electrons are the tail.  The short timescale for Coulomb
equilibration of these $\sim 2$ keV electrons with the cold population
would require that the cold population have an enormous density; it
would itself cool rapidly and produce copious unseen optical emission,
and the energy required to maintain the warmer population would be
exorbitant.  We then show that even the $\sim 5$ keV continuum near
the Fe K$\alpha$ line is unlikely to be nonthermal bremsstrahlung.
Again, the energetics of maintaining the hot population against energy
transfer into the cold population would require that a significant
portion of the blast energy go into electrons with energies of order 5
keV, which seems unreasonable.

We showed in \S\ref{ejecta} that the Fe K$\alpha$ line cannot be
produced in a thermal plasma with approximately solar abundances.  Is
it possible to account for the observed Fe K$\alpha$ emission by
postulating deviations from a Maxwellian electron distribution in such
plasma, as suggested by \citet{vink97,vink02}? First we consider
whether the featureless spectrum at lower energies ($\sim 2$ keV),
which we interpret as nonthermal synchrotron emission, could be
attributed to nonthermal bremsstrahlung produced by suprathermal
electrons accelerated in the shock wave. The following arguments based
on order-of-magnitude estimates demonstrate that this is not
possible. For the nonthermal bremsstrahlung hypothesis, one needs a
large thermal pool of cold electrons with temperature $T_c$, and much
less numerous suprathermal electrons forming a power-law tail at high
electron energies.  The description of these fast electrons in terms
of a power law is a crude approximation because of Coulomb losses to
the cold electrons, and because the particle acceleration process at
the shock may produce a curved (concave) spectrum instead of a power
law \citep{er91,re92}.  In view of these complexities, we do not
attempt to model the electron distribution function, but assume
instead a Maxwellian distribution of hot electrons. Such a
two-component (cold+hot) thermal model is sufficient for our present
purpose of estimating the number density of fast electrons which would
be required to account for the observed strength of the featureless
X-ray continuum. The bulk plasma temperature $T_c$ must be low enough
that the ionization state of the gas is so low that practically no
X-ray lines are produced in the {\it Chandra} energy range, to be
consistent with the observed featureless spectra. The observed X-ray
continuum would then be free-free emission produced by fast electrons
in collisions with protons and $\alpha$ particles. In order to
estimate the required density of fast electrons, we fit a standard
free-free emission model to the spectrum shown in Figure
\ref{fig:hard}. We obtain a hot electron temperature $T_h$ of 1.45
keV, and corresponding emission measure $EM_h = \int n_{eh} n_H dV$ of
$2.6 \times 10^{57} d_3^2$ cm$^{-3}$ (where $n_{eh}$ is the number
density of hot electrons, which must be much smaller than the hydrogen
density $n_H$). For the X-ray emitting volume of $37 d_3^3$ pc$^3$
estimated in \S\ref{ejecta}, $n_{eh} = 2.4/f_h n_H d_3$ cm$^{-3}$,
where $f_h$ is the unknown volume filling fraction of fast electrons.

The hot electron temperature of 1.45 keV implies an average energy $E$
for fast electrons of 2.2 keV. Electrons with such a low energy lose
it rapidly in Coulomb collisions with cold electrons in the thermal
pool. The Coulomb energy loss timescale $t_E$ is equal to $32
E^{3/2}/G((E/T_c)^{1/2}) n_e \ln \Lambda$ yr \citep{spitzer62}, where
$n_e = 1.2 n_H$ is the density of cold electrons, $\ln \Lambda$ is the
Coulomb logarithm, and the function $G$ is tabulated by
\citet{spitzer62}. Because expected electron densities are high ($n_e
\gg 1$ cm$^{-3}$), this is a short time scale. For example, for $E =
2.2$ keV and $T_c = 0.1$ keV, we find $\ln \Lambda = 30$,
$(E/T_c)^{1/2} = 4.7$, $G(4.7) = 0.024$, and $t_E = 140 n_e^{-1}$ yr.
This short Coulomb energy exchange time $t_E$ implies that 2 keV
electrons are strongly coupled to the low-temperature thermal plasma
through Coulomb collisions.  This strong thermal coupling demands that
the thermal energy in fast electrons be much smaller than in the cold
thermal pool, otherwise the energy so quickly transferred to the cold
electrons would raise their temperature enough to produce strong X-ray
lines, contrary to observations.

Let $R \equiv (n_e+n_H)T_c/n_{eh}T_h$ denote the cold/hot energy
ratio, which we demand to be much larger than unity. Using our
estimates for $n_{eh}$ obtained above from the strength of the X-ray
continuum, $R = 0.62 f_h n_e^2 d_3 T_c/T_h$. Because of the strong
coupling through Coulomb collisions, the volume filling fraction $f_h$
is expected to be low for shock-accelerated fast electrons, on the
order of $t_E/t_s$ or less, where $t_s$ is the shock age.  The
expected shock age $t_s$ should be a sizeable fraction of the
remnant's age, which is $\sim 10^4 d_3$ yr \citep{rosado96}.  Denoting
$t_{s,4}=t_s/10^4$ yr and using the expression for $t_E$ obtained above,
we arrive at $R = 0.002 E^{1/2} T_c n_e/t_{s,4}G((E/T_c)^{1/2}) \ln
\Lambda$.  By equating $R$ to 1, we obtain a conservative lower limit
to the cold electron density $n_e$ of $500 t_{s,4} G((E/T_c)^{1/2})
\ln\Lambda E^{-1/2} T_c^{-1}$ cm$^{-3}$.  This is a very large
electron density, e.~g., even for a relatively recent ($t_{s,4} =
0.2$) shock with $E = 2.2$ keV, and $T_c$ in the range of 0.05--0.1
keV, we arrive at $n_e \sim 500$ cm$^{-3}$.  Even if $T_c < 0.1$ keV,
the density of this cold plasma is large enough to produce a
significant amount of very soft X-ray emission which would be easily
seen in X-ray spectra.  In addition, such plasma is located close to
the peak of the cooling function, and it would itself cool and
recombine on timescales much smaller than the age of the remnant even
in the presence of magnetic pressure support.  Such a cooling and
recombining plasma should produce strong optical emission.  There is
no evidence for optical emission at the location of the hard X-ray
emitting filaments in the optical images presented by \citet{smith97}.
Another problem is the high ($6.5
\times 10^{-8}$ dynes cm$^{-2}$) pressure of hot electrons, which
exceeds the estimated ram pressure of the blast wave by an order of
magnitude or more. This problem becomes more acute if the filling
fraction $f_h$ is not small. In this case one must assume that an
additional energy source exists throughout the X-ray emitting plasma,
which could heat hot electrons and balance their Coulomb losses. The
power required to achieve this balance far outstrips the amount of
kinetic energy in the blast wave. We conclude that the observed X-ray
continuum emission at low energies cannot be produced by nonthermal
bremsstrahlung.

While the low-energy continuum is almost certainly synchrotron
radiation, the high-energy continuum in the vicinity of the Fe
$K\alpha$ line could in principle be produced by nonthermal
bremsstrahlung by high energy suprathermal electrons. Because of the
high ($\sim 5-10$ keV) electron energies required to produce this
continuum, such electrons will be less strongly coupled to the cold
electrons through Coulomb collisions than the $\sim 2$ keV electrons
considered above. The high energy continuum is nevertheless quite
strong, implying a high density $n_{eh}$ for these fast electrons and
nonnegligible thermal coupling between cold and fast electrons. 
We can repeat our previous arguments where now the subscript $h$ refers
to the $\sim 5 - 10$ keV electrons.  Using
our thermal fit to the high energy continuum and our estimates of the
X-ray emitting volume (see \S\ref{ejecta}), we arrive at $n_{eh} =
0.48/f_h n_H d_3$ cm$^{-3}$.  Next, we estimate the energy transfer
rate from hot to cold electrons, again approximating their electron
distribution functions by Maxwellians with low ($T_c$) and high
($T_h$) temperatures. This rate (per unit volume) is equal to
$-\case{3}{2}n_{eh}k dT_h/dt = \case{3}{2} n_{eh} k (T_h-T_c)/t_{eq}$,
where $t_{eq}$ is the equipartition time between two Maxwellian
electron distributions given by \citet{spitzer62}. With our estimate
for $n_{eh}$, and assuming that $T_c \ll T_h$, we arrive at the energy
transfer rate of $0.86kT_h/n_et_{eq}f_hd_3$. Note that because $t_{eq}
\propto n_e^{-1}$, this energy transfer rate is independent of the
density of cold electrons. For $kT_h = 5$ keV, $f_h = 1$, we arrive at
a conservative lower limit to the energy transfer rate of $3.4
\times 10^{-18} d_3^{-1}$ ergs cm$^{-3}$ s$^{-1}$. If we now neglect
fast electron energy losses, i.~e., if we assume that $T_h$ is
constant so that the fast electrons coexist with the cold electrons
throughout the whole X-ray emitting plasma, we find that in 1,000 yr
the total amount of transferred energy is equal to $1.1 \times
10^{-7}$ ergs cm$^{-3}$. This would result in a thermal pressure of
the cold plasma of $7.1 \times 10^{-8}$ dynes cm$^{-3}$, more than the
ram pressure of the blast wave, where velocities range up to 900 km
s$^{-1}$ and preshock densities may be as high as $\sim 1$ cm$^{-3}$
\citep{ghrs01}. While there are some uncertainties in estimates of the
preshock density because of the unknown preshock ionization fraction,
this finding suggests that $\sim 5 - 10$ keV electrons would play a
major role in the energy balance of the plasma. This is not expected
in standard particle acceleration models where most of the shock
kinetic energy is transfered to cosmic-ray particles, and would rule
out the simultaneous production of $\sim 100$ times as much energy in
accelerated ions, as observed in cosmic rays at Earth.

This problem remains if we relax our assumption of constant $T_h$ by
allowing for Coulomb energy losses. We assume that fast electrons are
injected at the shock front with energy $E$ and then lose their energy
on a timescale $t_E = 5.3 \times 10^3 n_e^{-1}$ yr ($t_E$ was
estimated here in the limit of $T_c = 0$). For expected electron
densities $n_e$ far in excess of 1 cm$^{-3}$, Coulomb energy losses
are important. The volume filling fraction $f_h$ for plasma containing
fast electrons is then small, less than $\sim t_E/t_s =
0.53t_{s,4}^{-1}n_e^{-1}$. With this volume fraction, we arrive at
$n_{eh} = 1.6 t_{s,4}$ cm$^{-3}$, and with $T_h = 5$ keV the pressure
of fast electrons is equal to $2.6 \times 10^{-9}$ dynes cm$^{-2}$
even for a relatively young ($t_{s,4}=0.2$) shock. This pressure is
substantial when compared with the estimated ram pressure, again
implying a major role in the plasma energetics for $\sim 5-10$ keV
electrons.  It is hard to explain this in the framework of our current
understanding of collisionless plasmas and the particle acceleration
process. While we cannot completely rule out nonthermal
bremsstrahlung as the explanation for the high-energy continuum on an
observational basis alone, we do not favor it because of excessive
energy requirements.

\clearpage

\def\res#1#2#3{$#1_{-#2}^{+#3}$}

\begin{deluxetable}{lccccccc}

\tablecolumns{8}

\tablewidth{0pc}

\tabletypesize{\footnotesize}

\tablecaption{Spectral Models \label{tab:models}}

\tablehead{

  \colhead{}  & \multicolumn{2}{c}{Soft Region} &

  \multicolumn{2}{c}{Hard Region (S3)}  &

  \multicolumn{2}{c}{Hard Filaments (S3)} & {I3 Filament} \\

  \colhead{Model Parameters} & A\tablenotemark{a} & B\tablenotemark{b} & A & B & A & B & A }

\startdata

$N_H/10^{21}$ cm$^{-2}$ & \res{5.88}{0.06}{0.03} & \res{5.26}{0.16}{0.16} & \res{4.43}{0.14}{0.14} & \res{4.16}{0.11}{0.10} & \res{4.94}{0.34}{0.35} & \res{4.61}{0.30}{0.41} & \res{3.80}{0.40}{0.37}\\

$kT$ (keV)  & \res{0.415}{0.008}{0.015} & \res{0.430}{0.017}{0.015} & \res{0.675}{0.057}{0.052} & \res{0.675}{0.101}{0.025} & \res{0.60}{0.17}{0.22} & \res{0.48}{0.13}{0.27} & \res{2.1}{1.1}{1.5}\\

$\tau/10^{10}$ cm$^{-3}$s & \res{2.77}{0.12}{0.10} & \res{5.09}{0.50}{0.67} & \res{1.74}{0.17}{0.18} & \res{2.34}{0.24}{0.77} & \res{1.71}{0.45}{0.96} & \res{3.88}{1.76}{5.16} & \res{2.20}{0.53}{2.59} \\

C, N       & \res{1.20}{0.08}{0.07} & \res{0.80}{0.17}{0.18} &1 & 0.80 &1 & 0.80 & 1 \\

O          & \res{1.20}{0.08}{0.07} & \res{0.86}{0.10}{0.11} &1 & 0.86 &1 & 0.86 & 1 \\

Ne         & \res{1.20}{0.08}{0.07} & \res{0.99}{0.09}{0.09} &1 & 0.99 &1 & 0.99 & 1 \\

Mg         & \res{1.20}{0.08}{0.07} & \res{0.67}{0.07}{0.06} &1 & 0.67 &1 & 0.67 & 1 \\

Si         & \res{1.20}{0.08}{0.07} & \res{0.92}{0.08}{0.09} &1 & 0.92 &1 & 0.92 & 1 \\

S          & \res{1.20}{0.08}{0.07} & \res{0.85}{0.19}{0.19} &1 & 0.85 &1 & 0.85 & 1 \\

Fe, Ni, Ca & \res{1.20}{0.08}{0.07} & \res{0.63}{0.06}{0.06} &1 & 0.63 &1 & 0.63 & 1 \\

$(EM/4\pi d^2)/10^{11}$ cm$^{-5}$  & \res{23.2}{1.7}{1.7} & \res{20.5}{1.8}{1.8} & \res{1.90}{0.30}{0.28} & \res{1.79}{0.20}{0.29} & \res{0.47}{0.16}{0.33} & \res{0.50}{0.20}{0.54} & \res{0.12}{0.03}{0.05} \\

$kT_{ej}$ (keV)  & \multicolumn{2}{c}{ } & 5 & 5 & 5 & 5 & \\

$\tau_{ej}$ (cm$^{-3}$s)   & \multicolumn{2}{c}{ } & $10^9$ & $10^9$ & $10^9$  

 & $10^9$ & \\

Fe K$\alpha$ $EW$ (keV) & \multicolumn{2}{c}{ } & \res{0.38}{0.13}{0.13} &

\res{0.37}{0.13}{0.13} & \res{0.17}{0.17}{0.17} & \res{0.16}{0.16}{0.18} & \\

$(EM_{ej}/4\pi d^2)/10^7$ cm$^{-5}$  & \multicolumn{2}{c}{ } & \res{2.91}{0.99}{0.99} & \res{2.88}{0.99}{0.99}

 & \res{0.26}{0.26}{0.26}& \res{0.24}{0.24}{0.27} & \\

$\nu_c/10^{16}$ Hz &  & & \res{8.08}{0.51}{0.51} & \res{8.58}{0.44}{0.40} & \res{7.68}{1.03}{1.10} & \res{8.19}{1.47}{0.89} & \res{10.1}{1.4}{2.0} \\

$\alpha$  &  &  & 0.6 & 0.6 & 0.6 & 0.6 & 0.6 \\

$F_{1 GHz}$ (Jy) &  &  & \res{2.80}{0.17}{0.17} & \res{2.59}{0.11}{0.15} & \res{0.61}{0.08}{0.09} & \res{0.57}{0.06}{0.11} & \res{0.48}{0.08}{0.07}\\

X\tablenotemark{c} (0.5-10 keV) & 0\% & 0\% & 83\% & 84\% & 86\% & 88\% & 86\%\\

Reduced $\chi^2$ & 5.87 & 2.69 & 1.43 & 1.39 & 1.05 & 1.03 & 1.16 \\

Counts/s (0.5-10 keV) & 2.5 & 2.5 & 2.0 & 2.0 & 0.37 & 0.37 & 0.27 \\

\enddata

\tablenotetext{a}{$pshock$}

\tablenotetext{b}{$vpshock$}

\tablenotetext{c}{Nonthermal/(nonthermal+thermal) flux ratio}

\end{deluxetable}

\clearpage

\begin{figure}

\caption{Total energy band (0.5--8 keV) {\it Chandra} image of
RCW 86. The scale ranges from $6 \times 10^{-8}$ to $1 \times 10^{-6}$
photons cm$^{-2}$ s$^{-1}$ arcsec$^{-2}$. The two chips in the left
column are, from top (north) down, I2 and I3; in the right column, 
S0, S1, S2, and S3. This and following images have been binned to 
a resolution of 2\arcsec.}
%\caption{0.5--10 keV {\it Chandra} image of RCW 86
%obtained with ACIS I2, I3, S0--S3 CCDs.}
\label{fig:totx}

\end{figure}

\begin{figure}

\caption{Mosaiced three-color {\it Chandra} images of RCW 86: red represents
0.5--1 keV photons; green, 1--2 keV; and blue 2--8 keV.}
\label{fig:truecolor}

\end{figure}

\begin{figure}

\caption{Soft (0.5--1 keV) X-ray image of RCW 86.
}

\label{fig:softx}

\end{figure}

\begin{figure}

\caption{Hard (2-10 keV) X-ray image of RCW 86.}

\label{fig:hardx}

\end{figure}

\begin{figure}

\caption{H$\alpha$ image from \citet{smith97} (in green) compared
with the soft (0.5--1 keV) X-ray image (in red).}

\label{fig:Halpha}

\end{figure}

\begin{figure}

\caption{The 1.4 GHz ATCA image \citep{dsm01} (in red) compared with
hard (2-10 keV) and soft (0.5--1 keV) {\it Chandra} images (in blue
and green, respectively).}  The radio image has been smoothed to a 
resolution of $8^{\prime\prime}$.

\label{fig:xrayradioc}

\end{figure}

\begin{figure}

%\plotone{regprof.eps}

\caption{{\it Chandra} image in the 0.5--8 keV energy band
with spectral extraction regions and spatial profiles  
indicated (in red and yellow, respectively).}

\label{fig:extract}

\end{figure}

\begin{figure}

%\plotone{figprof.eps}

\caption{Spatial profiles of hard (2--8 keV) X-ray emission and
1.4 GHz radio emission ({\it solid} and {\it dashed} lines,
respectively) for three positions shown in 
Figure~\ref{fig:extract}, from west ({\it top}) to east 
({\it bottom}).
X-ray emission is scaled independently for each
profile relative to radio emission. 
Distance is measured in arcseconds, from E to W
({\it top} profile), and from NE to SW ({\it middle} and {\it bottom}
profiles). }

\label{fig:profiles}

\end{figure}

\begin{figure}

%\plotone{soft.ps}

\caption{ACIS S3 X-ray spectrum of the bright nonradiative
Balmer filament. Plane-parallel shock model (model B from
Table \ref{tab:models}) is shown by 
{\it solid line}. }

\label{fig:soft}

\end{figure}

\begin{figure}

%\epsscale{0.9}

%\plotone{hard.ps}

\caption{ACIS S3 X-ray spectrum of the entire synchrotron-emitting
region. Model B from Table \ref{tab:models} is shown by {\it solid line}.
Individual models components are also shown: {\it srcut -- short-dashed line},
pure Fe {\it vpshock} model -- {\it long-dashed}, low-temperature
{\it vpshock} model  -- {\it dash-dotted}.
}

\label{fig:hard}

\end{figure}

\begin{figure}

%\epsscale{1.0}

%\plotone{synch.ps}

\caption{ACIS S3 (top) and I3 (bottom) X-ray spectra of 
synchrotron-emitting filaments, with best-fit models from
Table \ref{tab:models} shown by {\it solid} lines. Individual
model components are also shown. }

\label{fig:synchrotronI}

\end{figure}

\clearpage

%\plotone{f1color.ps}

%{Fig. 1}

%\psfig{figure=f2color.ps,width=20truecm}  

%\plotone{f2color.ps}

%{Fig. 2}

%\psfig{figure=f3.ps,width=16truecm} 

%\plotone{f3.ps}

%{Fig. 3}

%\psfig{figure=f4.ps,width=16truecm} 

%\plotone{f4.ps}

%{Fig. 4}

%\psfig{figure=f5color.ps,width=19truecm} 

%\plotone{f5color.ps}

%{Fig. 5}

%\psfig{figure=f6color.ps,width=19truecm} 

%\plotone{f6color.ps}

%{Fig. 6}

%\psfig{figure=f7color.eps}

\plotone{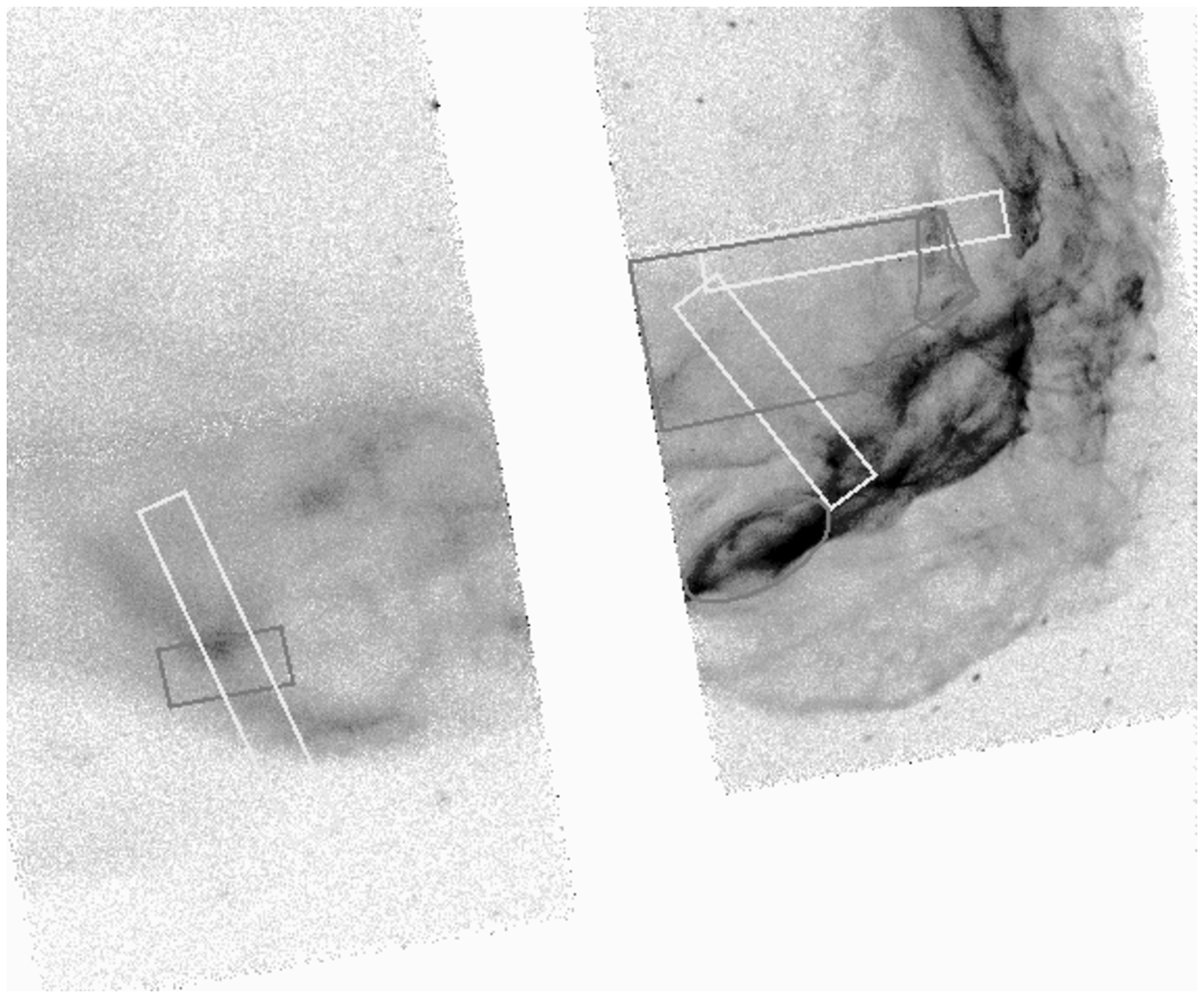}

{Fig. 7}

\plotone{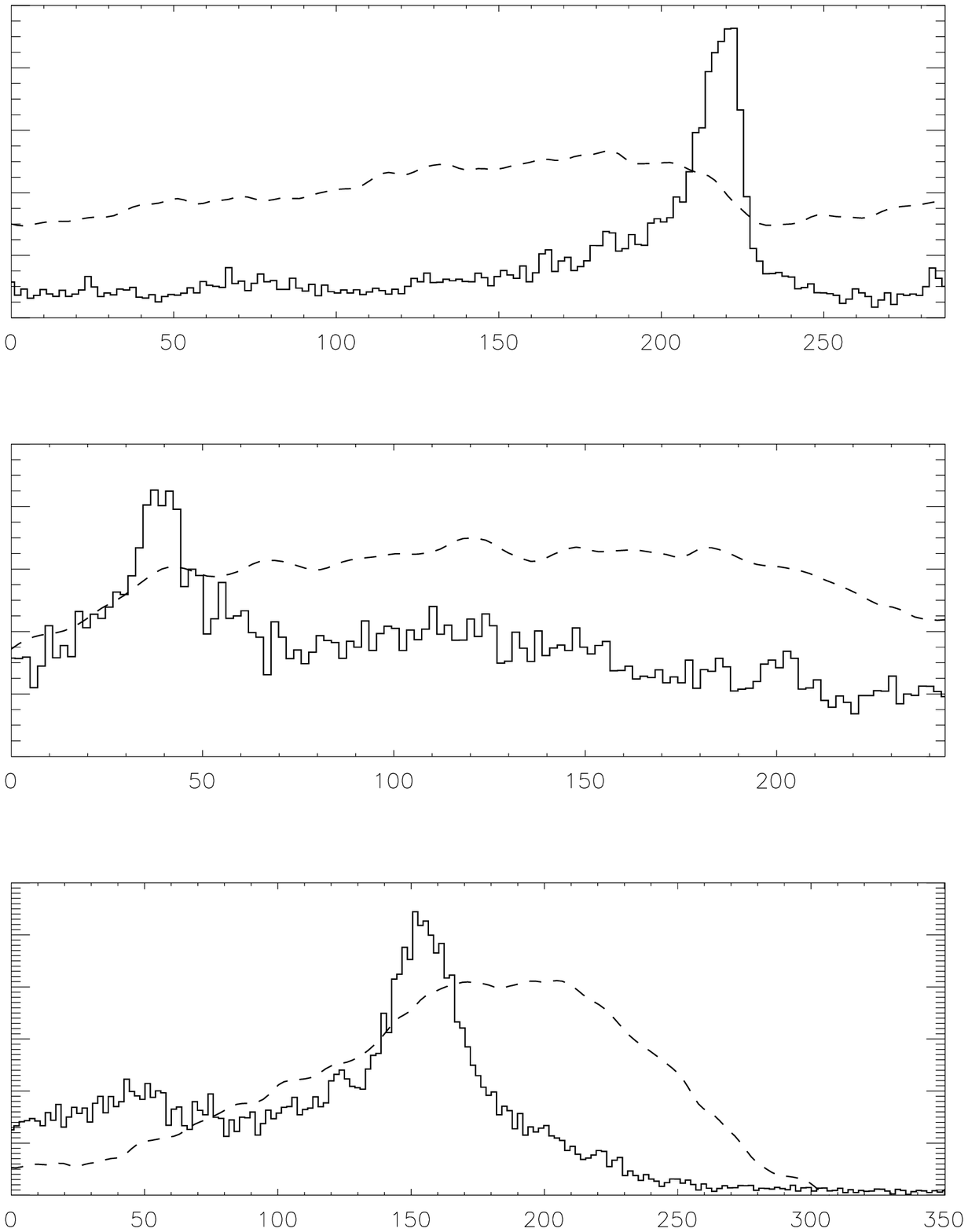} 

{Fig. 8}

\plotone{f9.ps}

%{Fig. 9}

%\psfig{figure=f10.ps,height=8truecm,width=12truecm}

\plotone{f10.ps}

%{Fig. 10}

%\psfig{figure=f11.ps}

\plotone{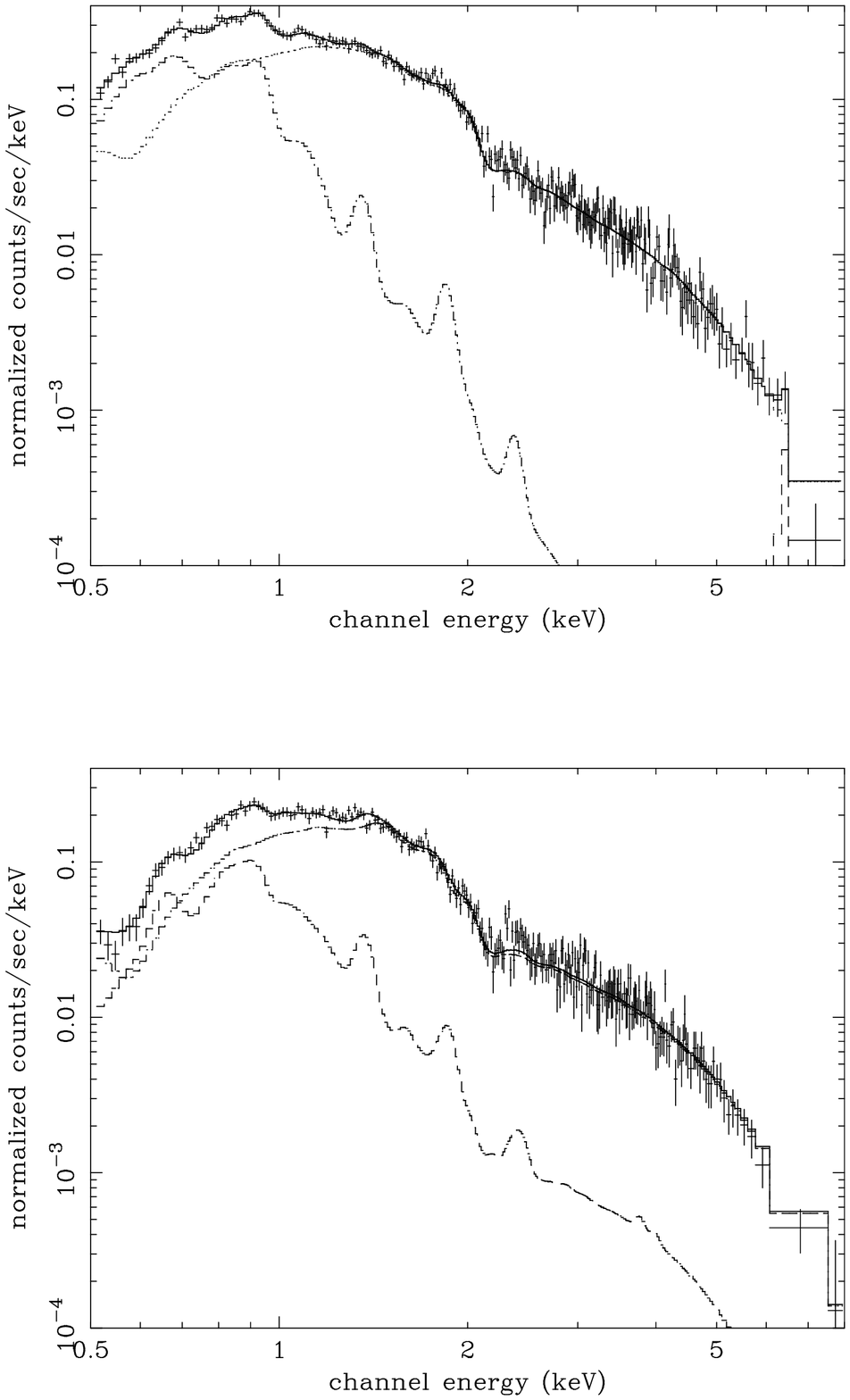}

%{Fig. 11}


\begin{thebibliography}{}        



\bibitem[Allen et al.(1999)]
{allen99}
Allen, G.~E., Gotthelf, E.~V., \& Petre, R. 1999, in the Proceedings of the
26th International Cosmic Ray Conference, Salt Lake City

\bibitem[Arnaud(1996)]
{arnaud96}
Arnaud, K.~A. 1996, in Astronomical Data Analysis and Systems V, eds. 
G.Jacoby \& J.Barnes, ASP Conf. Series, v.101, 17



\bibitem[Bamba, Koyama, \& Tomida(2000)]
{bamba00}
Bamba, A., Koyama, K., \& Tomida, H. 2000, PASJ, 52, 1157

\bibitem[Bamba et al.(2001)]
{bamba01}
Bamba, A., Ueno, M., Koyama, K., \& Yamauchi, S. 2001, PASJ, 53, L21


\bibitem[Bocchino et al.(2000)]
{bvfms00}
Bocchino, F., Vink, J., Favata, F., Maggio, A., \& Sciortino, S. 2000, A\&A,
360, 671



\bibitem[Borkowski, Lyerly, \& Reynolds(2001)]
{blr01}
Borkowski, K. J., Lyerly, W. J., \& Reynolds, S. P. 2001, ApJ, 548, 820



\bibitem[Borkowski et al.(2001)]
{brrd01}
Borkowski, K. J., Rho, J.,  Reynolds, S. P., \& Dyer, K. K. 2001, ApJ, 550, 334



\bibitem[Brickhouse et al.(2000)]
{bretal00}
Brickhouse, N. S., Dupree, A. K., Edgar, R. J., Liedahl, D. A., Drake, S. A.,
White, N. E., \& Singh, K. P. 2000, ApJ, 530, 387



\bibitem[Dickel et al.(2001)Dickel, Strom, \& Milne]
{dsm01}
Dickel, J. R., Strom, R. G., \& Milne, D. K. 2001, ApJ, 546, 447



\bibitem[Dyer et al.(2001)]
{dyer01}
Dyer, K. K., Reynolds, S. P., Borkowski, K. J., Allen, G. E., \&  Petre, R.
2001, ApJ, 551, 439



\bibitem[Ellison, Jones, \& Ramaty(1989)]
{ejr89}
Ellison, D., Jones, C. F., \& Ramaty, R. 1989, 21st International Cosmic Ray 
Conference, 4, 68



\bibitem[Ellison \& Reynolds(1991)]
{er91}Ellison, D., \& Reynolds, S. P. 1991, ApJ, 382, 242



\bibitem[Garmire et al.(1992)]
{garmire92}
Garmire, G. P., et al. 1992, in Proc AIAA, Space Programs and Technologies Conference



\bibitem[Ghavamian et al.(2001)]
{ghrs01}
Ghavamian, P., Raymond, J., Smith, R.~C., \& Hartigan, P. 2001, ApJ, 547, 995



\bibitem[Green(2001)]
{green01}
Green, D.~A. 2001, A Catalogue of Galactic Supernova Remnants (2001
December version), Mullard Radio Astronomy Observatory, Cavendish
Laboratory, Cambridge, United Kingdom 




\bibitem[Hamilton \&\ Sarazin(1984a)]
{hamsar84a}
Hamilton, A.~J.~S., \&\ Sarazin, C.~L. 1984a, ApJ, 284, 601



\bibitem[Hamilton \&\ Sarazin(1984b)]
{hamsar84b}
Hamilton, A.~J.~S., \&\ Sarazin, C.~L. 1984b, ApJ, 287, 282



\bibitem[Hamilton, Sarazin, \& Szymkowiak(1986)]
{hss86}
Hamilton, A.~J.~S., \&\ Sarazin, C.~L., \& Szymkowiak, A. E. 1986, 
ApJ, 300, 698



\bibitem[Jokipii(1987)]
{jokipii87}
Jokipii, J. R. 1987, ApJ, 313, 842



\bibitem[Koyama et al.(1997)]
{koyama97}
Koyama, K., et al. 1997, PASJ, 49, L7



\bibitem[Koyama et al.(1995)]
{Koyama95}
Koyama, K., Petre, R., Gotthelf, E. V., Hwang, U., Matsuura, M.,
Ozaki, M., \& Holt, S. S. 1995, Nature 378, 255



\bibitem[Laming(1998)]
{laming98}
Laming, J. M. 1998, ApJ, 499, 309



\bibitem[Levinson(1992)]
{levinson92}
Levinson, A. 1992, ApJ, 401, L73



\bibitem[Levinson(1994)]
{levinson94}
Levinson, A. 1994, ApJ, 426, L327



\bibitem[Liedahl et al.(1995)]
{liedahl}
Liedahl, D.~A., Osterheld, A.~L., \&\ Goldstein, W.~H. 1995, ApJ, 438, L115



\bibitem[Long \& Blair(1990)]
{lobl90}
Long, K.~S., \& Blair, W.~P. 1990, ApJ, 358, L13



\bibitem[Ramaty \& Lingenfelter(1979)]
{rl79}
Ramaty, R., \& Lingenfelter, R.E. 1979, 16th International Cosmic Ray 
Conference, 1, 501 



\bibitem[Reynolds(1998)]
{R98} 
Reynolds, S. P. 1998, \apj, 493, 375



\bibitem[Reynolds \& Ellison(1992)]
{re92}
Reynolds, S. P., \& Ellison, D. 1992, ApJ, 399, L75



\bibitem[Reynolds \& Keohane(1999)]{rk99} Reynolds, S.P., \&
Keohane, J. W. 1999, ApJ, 525, 368

\bibitem[Rosado et al.(1996)]
{rosado96}
Rosado, M., Ambrocio-Cruz, P., Le Coarer, E., \& Marcelin, M. 1996, A\&A, 315, 243



\bibitem[Slane et al.(1999)]
{slane99} 
Slane, P. , Gaensler, B.  M.,
Dame, T. M., Hughes, J. P., Plucinsky, P. P. \& Green, A.  1999, \apj,
525, 357



\bibitem[Slane et al.(2001)]
{slane01} 
Slane, P., et al. 2001, ApJ, 548, 814



\bibitem[Smith(1997)]
{smith97}
Smith, R.~C. 1997, AJ, 114, 2664



\bibitem[Spitzer(1962)]
{spitzer62}
Spitzer, L. 1962, Physics of Fully Ionized Gases (New York: Wiley)



\bibitem[Townsley et al.(2000)]
{townsley00}
Townsley, L. K., Broos, P. S., Garmire, G. P., \& Nousek, J. A. 
2000, ApJ, 534, L139



\bibitem[Vink et al.(2002)]
{vink02}
Vink, J., Bleeker, J.~A.~M., Kaastra, J.~S., van der Heyden, K.~J.,
Rasmussen, A.~P., \& Dickel, J.~R. 2002, in New Visions of the
X-ray Universe in the {\it XMM-Newton} and {\it Chandra} Era, ESTEC, in press



\bibitem[Vink et al.(1997)]
{vink97}
Vink, J., Kaastra, J.~S., \& Bleeker, J.~A.~M. 1997, A\&A, 328, 628



\bibitem[Weisskopf, O'Dell, \& van Speybroeck(1996)]
{weiss96}
Weisskopf, M. C., O'dell, S. L., \& van Speybroeck, L. P. 1996, 
Proc. SPIE, 2805, 2



\bibitem[Whiteoak \& Green(1996)]
{wg96}
Whiteoak, J.~B.~Z., \& Green, A. J. 1996, A\&AS, 118, 329 



\end{thebibliography}
\end{document}